\documentclass[a4paper,twoside]{article}

\usepackage{epsfig}
\usepackage{subfigure}
\usepackage{calc}
\usepackage{amssymb}
\usepackage{amstext}
\usepackage{amsmath}
\usepackage{amsthm}
\usepackage{multicol}
\usepackage{pslatex}
\usepackage{apalike}
\usepackage{SCITEPRESS}
\usepackage[small]{caption}

\usepackage[T1]{fontenc}
\usepackage{ifxetex,ifluatex}
\ifnum 0\ifxetex 1\fi\ifluatex 1\fi=0 % if pdflatex
  \usepackage[utf8]{inputenc}
\fi

\makeatletter
\def\maxwidth{\ifdim\Gin@nat@width>\linewidth\linewidth
\else\Gin@nat@width\fi}
\makeatother
\let\Oldincludegraphics\includegraphics
\renewcommand{\includegraphics}[1]{\Oldincludegraphics[width=\maxwidth]{#1}}

\def \citep {\cite}

\begin{document}

\title{Review of the Use of Electroencephalography as an Evaluation Method for Human-Computer Interaction}

\author{\authorname{Jérémy Frey\sup{1,2,3}, Christian Mühl \sup{3}, Fabien Lotte \sup{3} and Martin Hachet \sup{3}}
\affiliation{\sup{1}Univ. Bordeaux, LaBRI, UMR 5800, F-33400 Talence, France}
\affiliation{\sup{2}CNRS, LaBRI, UMR 5800, F-33400 Talence, France}
\affiliation{\sup{3}INRIA, F-33400 Talence, France}
\email{jeremy.frey@inria.fr, c.muehl@gmail.com, fabien.lotte@inria.fr, martin.hachet@inria.fr}
}

\keywords{\textsc{HCI evaluation, EEG, ErrP, workload, attention, emotions}}

\abstract{Evaluating human-computer interaction is essential as a broadening population uses machines, sometimes in sensitive contexts. However, traditional evaluation methods may fail to combine real-time measures, an ``objective'' approach and data contextualization. In this review we look at how adding neuroimaging techniques can respond to such needs. We focus on electroencephalography (EEG), as it could be handled effectively during a dedicated evaluation phase. We identify workload, attention, vigilance, fatigue, error recognition, emotions, engagement, flow and immersion as being recognizable by EEG. We find that workload, attention and emotions assessments would benefit the most from EEG. Moreover, we advocate to study further error recognition through neuroimaging to enhance usability and increase user experience.}

\onecolumn

\maketitle

\normalsize

\vfill

\section{INTRODUCTION}

\noindent Along computer science history, interfaces and interactions
have been getting more complex. Nowadays computers are everywhere, used
by everyone. It is necessary to make them comply with human
capabilities, practical to use. This is mostly done by evaluating HCI
prior to their public availability. Yet traditional evaluation methods
could either be ambiguous, lack real-time recordings, or disrupt the
interaction.

On the other hand, new technologies arise. Physiological sensors help to
improve the ergonomics of human-computer interaction (HCI)
\citep{Fairclough2009a}. Systems could be tuned to users by monitoring
their mental workload in real-time \citep{Kohlmorgen2007}. Physiological
sensors add an insightful information channel. However sensors may be
intrusive or require a calibration to record a proper signal, and some
are hardly available to consumers.

These issues could be resolved by using physiological sensors in HCI
evaluation. While designing a user interface (UI) it should be
acceptable to add sensors' hindrance to specially enrolled users. Those
testers will then help to improve beforehand the UI. Laboratory
conditions permit a controlled setup for repeatable measures.
Neuroimaging rely on demanding but sensitive sensors. We consider them
as an innovative supplement to conventional evaluation methods.
Measuring neural activity during HCI can help us to better understand
what occurs in the brain when users are interacting
\citep{Parasuraman2013}.

We highlight in this paper which neuroimaging techniques could be used
conveniently within laboratories to overcome the difficulties
encountered by traditional evaluation methods alone. We review a
repertoire of patterns of users' state which could be used to
characterize HCI, and evaluate how neuroimaging objectively measures
them. We call those patterns ``constructs'', a term which refers to
notions as different as workload and the state of ``flow''.

Other papers already began to sense how neurotechnologies benefit HCI,
but they do not cover evaluation \citep{George2010}, or if so they do
not study many constructs. \citep{Parasuraman2013} only discuss
workload, vigilance and error recognition. In the present review we
gathered from the HCI literature every major construct which could
potentially be evaluated with brain activity.

In this review, we first briefly describe the different families of
evaluation methods aimed at assessing HCI and UI quality, along with
their advantages and drawbacks. We divided them in four categories:
behavioral studies (observations of users actions in real-time),
inquiries (e.g.~questionnaires, interviews, think aloud), physiological
sensors (e.g.~heart rate, galvanic skin response) and neuroimaging (a
subset of physiological sensors which records brain activity). We also
formalize a new scale (whenever the measure is ``exocentic'' or
``egocentric'') which could help to choose the right combination of
methods for evaluations.

We show that electroencephalography (EEG) is the neuroimaging technique
which offers the best trade-off between spatial and temporal resolution,
practical use and cost. Therefore we focus on EEG during the second
part. We review there constructs related to the quality of HCI. We
identified that workload, attention, vigilance, fatigue, error
recognition, emotions, engagement, flow and immersion are useful for
evaluation and can be measured with EEG.

Finally we outline the challenges and limitations which arise from this
encounter between HCI evaluation and neurotechnologies, as well as
constructs that could benefit from being measurable with EEG.

\section{EVALUATION METHODS}

\subsection{Behavioral Studies}

Recording users interactions, such as mouse speed, is one standard way
to evaluate a UI. ``Behavioral studies'' refers to this method: behavior
and actions of users inside a software. Behavioral studies are close to
performance measures, as seen in human factors. The easiest way to sense
if a UI is well designed is to watch users. How fast do they complete
the task? Are they more accurate with a bigger mouse cursor? Such
methods helped to formulate a preeminent law in HCI, Fitts's law, which
is all about time to reach a target depending on its distance and size
\citep{Fitts1954}.

Although behavioral studies are able to account in real-time for users'
interactions, they can be hard to interpret: measures may not be
specific to one construct. E.g. a high reaction time can be caused
either by a low concentration level or a high workload
\citep{Berka2007}, \citep{Hart1988}. On top of that, behavioral studies
may not provide much information on the users' state. With simple tasks
in particular, little can be computed beside reaction times and a
performance metric.

\subsection{Inquiries}

\label{inquiries}

While it is possible to infer users' thoughts through a behavioral
study, it may be simpler to record their opinion. We call this
``inquiries''. In HCI we are interested in questionnaires related to the
use of a UI. Standardized questionnaires have been validated across
several studies for various measures: e.g.~NASA-TLX for workload
\citep{Hart1988}.

Unfortunately those ``pen and paper'' tests are discrete and are not
good for real-time assessments. The ``think aloud'' protocol
\citep{Weber2007} is a way to circumvent this, yet it could influence
the interaction as users still have two different things to do: interact
and report their experience. It is an example of double task and divided
attention \citep{Ogolla2011}. ``Focus groups'' \citep{Bruseberg2002} is
the third form of inquiry. It involves experts and advanced users, who
exchange about their findings under the control of the designer.

Questionnaires, think aloud an focus group are three different forms of
inquiry fraught with the same hazards. Resulting measures are prone to
be contaminated by ambiguities \citep{Nisbett1977}, social pressure
\citep{Picard1995} or participants' memory limitations
\citep{Kivikangas2010} -- when answers are not oriented toward
experimenters' expectations if subjects figure out what is at stake.

\subsection{Physiological Sensors}

When humans interact with computers bodily changes co-occurs with mental
changes. E.g. pupils dilate while experiencing strong emotions
\citep{Partala2003}. Physiological sensors can be used in order to
account for such body changes in HCI \citep{Fairclough2009a},
\citep{Dirican2011} or game \citep{Ravaja2009}, \citep{Nacke2009a}
research. Galvanic skin response (GSR, also called ``electrodermal
activity'') is among those sensors, as well as electrocardiography (ECG,
the signal modality heart rate is derived from) and electromyography
(EMG, caused by muscular activity, including facial expressions).

Even if someone trained could control his heartbeat, physiological cues
are great for the ``objectivity'' they bring into HCI (see section
\ref{exoego}). Body reactions are sometimes misleading though: you may
record ECG to study attention, whereas an increase in heartbeat can also
be caused by strong feelings. Muscles and organs are controlled by the
peripheral nervous system. Physiological sensors are a second-order
inference about the processing which occurs in the central nervous
system.

\subsection{Neuroimaging}

Neuroimaging is a currently rising field used in brain-computer
interfaces (BCI) settings \citep{Blankertz2010},
\citep{Hamadicharef2010}. Neuroimaging techniques allows the assessment
of brain activity; we classify them apart even if strictly speaking they
do belong to physiological sensors.

Non-invasive neuroimaging techniques, which do not require surgery, are
divided into two main families \citep{Zander2011}. Functional magnetic
resonance imaging (fMRI) and functional near-infrared spectroscopy
(fNIRS) record brain activity through blood flow variations. fMRI has a
very good spatial resolution but is a large device which completely
surrounds subjects and costs about one million dollars. fNIRS is a much
more lightweight and affordable device. Instead of magnetic fields, it
uses direct light for recordings. Sensors are fixed on a cap, hence
subjects are free to interact with a computer while wearing it. Compared
to fMRI, the spatial resolution of fNIRS is less detailed. It records
only the outer region of the brain -- light is absorbed by tissues. fMRI
and fNIRS share a poor temporal resolution. With a latency reaching up
to several seconds it is difficult to observe fast and short responses.

The second family of neuroimaging uses electrical currents generated by
neural activity. Magnetoencephalography (MEG) records magnetic fields.
It is less heavy and expensive than fMRI, but still hardly manageable
for uses in HCI contexts. MEG has a high temporal resolution, down to
the millisecond. Electroencephalography (EEG) also has a high temporal
resolution. It is comparable in size to fNIRS. EEG measures electrical
current onto the scalp. Electrodes are ``dry'' -- no electrolyte
solution -- or, more frequently, ``wet'' -- solvent is either water or
gel. Despite its poor spatial resolution it is a relatively cheap
equipment for a laboratory. Because it is portable and non invasive, it
interferes little with HCI setting.

Experimenters must be cautious with the limitations of the device they
choose. Is the signal-to-noise ratio sufficient for what they intend to
measure? What artifacts could pollute their data? Are they in control of
the algorithms producing measures from raw signals? That said, EEG is
the most promising candidate to assist inquiries and other physiological
sensors in a wide range of evaluation measures. Compared to others
neuroimaging devices, EEG offers the best compromise between spatial and
temporal resolution, practical use and cost. Therefore we focus mostly
on this type of brain activity recordings in this paper.

\subsection{A New Continuum for HCI Evaluation Methods}

\label{exoego}

We have previously mentioned how the evaluation methods do bring
different levels of ``objectivity'' in their measures. Unfortunately, in
such context ``objective'' and ``subjective'' are scarcely defined in
the literature. According to \citep{Laar2013}, ``the objective methods
are based on overt and covert user responses during interaction while
the subjective methods rely on user expressions after the interaction''.
From that perspective, inquiries are ``subjective'' while behavioral
studies, physiological sensors and neuroimaging are ``objective''.

While we agree such a distinction is required, a more rigorous
vocabulary is needed. We also doubt the ``time'' variable should be
involved in the definition. As stated in section \ref{inquiries},
results of inquiries are prone to social pressure and other
self-interpretations, and this is also true for the real-time think
aloud. Moreover, when studying emotions, it could be argued that only
``subjective'' feelings are recorded, as the evaluation is centered on
the user. Hence, without a complex phrasing (i.e. ``objective measure of
subjective feelings''), employing such words is open to criticisms. As
an alternative ``direct'' and ``indirect'' could be considered. But then
those concepts are more likely to refer to how measures are reported,
not where they originate from (e.g.~EMG vs an external observer
annotating facial expressions).

As such, we would like to introduce a new nomenclature to name those two
aspects and avoid ambiguities: exocentric and egocentric. Those terms
are borrowed from spacial navigation research \citep{Brandt1973} and
bring the notion of the self. Exocentric measures are here close to the
stimuli, to the source, while egocentric measures are close to the
conscious thoughts of the user, to the outcome.

\begin{figure}[htbp]
\centering
\includegraphics{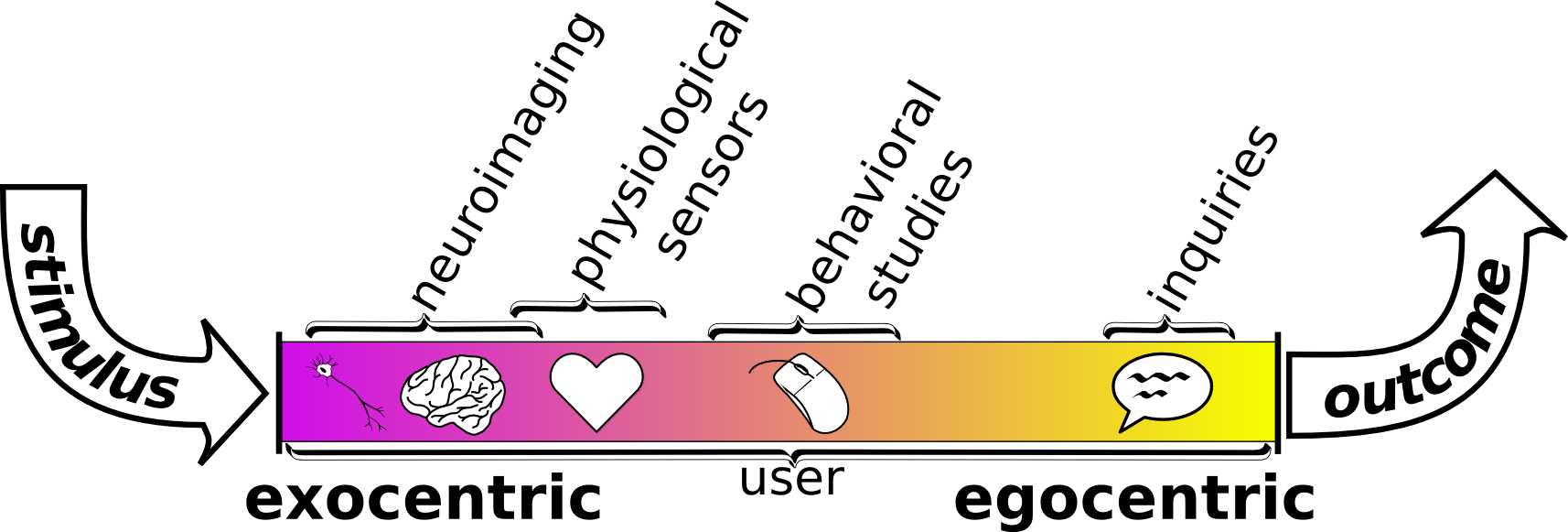}
\caption{Proposal of an ``exocentric / egocentric'' scale aimed at
classifying evaluation methods for HCI. \label{exo_ego_fig}}
\end{figure}

We therefore create a continuous space between two extremes (see Figure
\ref{exo_ego_fig}). We illustrate this scale with the measurement of
pain. The pressure of a needle on a finger would represent a perfect
exocentric measure: the stimulus' strength, a value disconnected from
human body and perceptions. When the pressure is transmitted to
nociceptors in the skin, the measure shifts a little from exocentric to
egocentric. As nerves are transmitting signals from the peripheral
nervous system to the brain, we go further to the right of the axis.
Since we may not be interested in skin's thickness, this neural activity
represents the first interesting value from this side of the
exo/egocentric scale. Neuroimaging techniques record such activity,
hence it is the most exocentric evaluation method. When the signal
reaches the central nervous system, autonomic responses are triggered --
increase in heart rate, galvanic skin response \citep{Loggia2011}. Those
reactions could be recorded through physiological sensors, a step
further from the exocentric extreme.

As the pain grows, it will alter behaviors and thoughts. A runner may
slow down when experiencing pain in a foot, no matter his willingness.
Behavioral studies are able to sense modifications occurring against the
will of the subject; that could be placed somewhere in the middle of our
scale. Concurrently, most of the time, the person is being aware of the
pain and could phrase it if asked to. Many other cognitive processes are
involved in such a high level of consciousness (e.g.~planning,
awareness), thus measures recorded by inquiries are close to the far-end
of the scale and are indeed egocentric.

This scale can be used for various evaluations. Eventually, it is
possible to add ``objective/subjective'' and ``direct/indirect'' to
describe a whole framework. A construct could be objective (usability)
or subjective (emotions). A tool could be either direct (sensor) or
indirect (observer). A method is more exocentric (neuroimaging) or
egocentric (inquiries). E.g. the work of an experimenter assessing
workload with ECG can be described as objective/exocentric/direct.

\section{CONSTRUCTS}

\noindent ``Constructs'' designate the patterns of users' state which
could be used to characterize interactions. This part reviews relevant
constructs from an HCI evaluation perspective that can be assessed using
neuroimaging techniques. We grouped similar measurements.

\subsection{Workload}

\subsubsection{Definition}

Humans have a limited set of resources to process information
\citep{Just2003}. The ratio between processing power and data coming
from the environment determines mental workload. Workload increases as
cognitive resources lessen or as the quantity of demands grows. If the
workload is too high subject's performance decreases, sometimes
dramatically.

\subsubsection{Neuroimaging}

Using a device with 9 channels \citep{Berka2007} correlated EEG with
workload. With a better equipment \citep{Mathan2007} showed how EEG
measures more subtle changes compared to ECG. fNIRS is another
well-tried technology: neurons require more energy, hence more oxygen,
as the load increases. fNIRS showed better results compared to EEG, with
82\% of correct classifications between 2 classes (low vs high workload)
and 50\% with 3 classes (low, medium, high) \citep{Hirshfield2009}.

In \citep{Blankertz2010} EEG online analyses (i.e.~in real-time)
discriminate 2 classes with a 70\% accuracy. A 2 minutes time window
enables scores from 80\% to 90\% \citep{Brouwer2012}. With 2 classes
still, reviews report scores close to 100\% if EEG is combined with
other physiological sensors \citep{Erp2010}. \citep{Grimes2008} claim
99\% success in distinguishing 2 memory load levels, 88\% with 4.

\subsection{Attention -- Vigilance -- Fatigue}

\subsubsection{Definition}

Attention, vigilance and fatigue are closely related and regularly
measured altogether \citep{Oken2006}.

``Attention'' refers to the ability to focus cognitive resources on a
particular stimulus \citep{Kivikangas2010}. A correct selective
attention allows to ignore distractors. An insufficient attention level
results in a difficulty or an inability to complete the task, whereas
too high or narrow attention resources may prevent someone to disengage
from a sub-task.

While in the literature, ``attention'' designates more frequently the
ability to perceive changes from the environment, the term ``vigilance''
then often refers to a broader resource, dependent of both cognitive
performance and the arousal level on the sleep--wake spectrum
\citep{Oken2006}. In that sense it refers to a state of sustained
attention. One needs to maintain a high degree of vigilance over time in
order to focus his attention on something. Hereby ``alertness'' will be
considered as a synonym of ``vigilance''.

``Fatigue'' is a state in which cognitive resources are exhausted. If
the required level of vigilance or attention causes a strain too
important on the organism, fatigue arises and performances decrease
\citep{Boksem2005}. Then the task cannot be performed correctly and
errors appear \citep{Erp2010}.

\subsubsection{Neuroimaging}

The alpha band is associated with attention. When eyes are closed, or
when fatigue occurs, alpha waves amplitude increases \citep{Shaw2003}.
This frequency band in the range 8-12Hz is mostly generated by the
occipital lobe. It is easily recorded with EEG, even with a single
electrode \citep{George2011}. Alpha band analysis discriminates
different attention levels \citep{Klimesch1998}. Even more, it enables
to detect which side of his visual field a subject is paying attention
to while his eyes stare in front of him with 70\% accuracy
\citep{Trachel2013}.

Other types of brain activity are used, such as delays in event-related
potentials (ERP) -- e.g.~visual selective attention in
\citep{Saavedra2012}.

\citep{Berka2007} suggested that EEG is the only sensor which can
accurately report attention and vigilance shifts on a second-by-second
timeframe. Works investigating vigilance measures are reviewed in
\citep{Parasuraman2013}.

Regarding fatigue, if EEG signals are not more accurate than
physiological sensors to detect microsleeps, they offer the possibility
to detect preceding inattentive states \citep[ sec.~3.1]{Blankertz2010}.
Mental fatigue has been detected on 4 seconds time windows with 80\%
accuracy, or 94\% over 30 seconds \citep{Laurent2013}. In order to
improve reliability, additional frequency ranges were recorded in this
study. For instance alpha, theta (4-8Hz) and beta (13-18Hz) bands have
been combined. ERP on the other hand have been used to study how fatigue
impairs differently cognitive processes \citep{Lorist2000}.

\subsection{Error Recognition}

\subsubsection{Definition}

We call ``error recognition'' the situation that occurs when users
detect by themselves an outcome different from what is expected
\citep{Nieuwenhuis2001}. It can be something users genuinely trigger but
then they realize they did a mistake. Or it can happen due to commands
erroneously interpreted by the machine.

Error recognition does not occur when a negative feedback is given per
se \citep{Ferrez2008}. It is a matter of recognition by the user of a
faulty event. In UI evaluation, error recognition could be an objective
measure of subjective (mis)representations, an objective assessment of
how intuitive an HCI is.

\subsubsection{Neuroimaging}

ERP are ``peaks'' and ``valleys'' in averaged EEG recordings associated
with an external event. ERP differ in their ``shapes'', place on the
scalp and latency depending on the source of the stimuli or on the
underlying cognitive mechanism. One particular kind of ERP has been
discovered: error-related potentials (ErrP) \citep{Schalk2000}. They are
triggered when an ``error'' occurs. It can be caused by something users
themselves did (response ErrP), by an incorrect response from the
command they used (interaction ErrP), by something they witnessed from
another user (observation ErrP), and also when an explicit negative
feedback is given (feedback ErrP). All of which have distinguishable
features \citep{Ferrez2008}.

Response ErrP and interaction ErrP suit perfectly our definition of
``error recognition''. Brain signals are elicited even when users are
not consciously aware of errors \citep{Nieuwenhuis2001}. ErrP have been
used to discriminate between incorrect and correct users decisions. In
\citep{Chavarriaga2010a} respectively 76\% and 63\% accuracy were
obtained to detect observation ErrP in ``single trial'', i.e.~in
detecting ErrP for each user's action.

These scores are common in the literature: 79\% and 84\% in a task
involving interaction ErrP \citep{Ferrez2008}. Accuracy relates to EEG
devices' quality. From 70\% with an entry-level headset and non
gel-based electrodes \citep{Vi2012} up to 90\% with a more expansive
device \citep{Schmidt2012}. While ErrP detection does not reach 100\%
(chance is 50\%), those scores are sufficient to improve HCI reliability
\citep{Vi2012}.

\citep{Sobolewski2013} recorded EEG while subjects use a mouse and have
to reach different targets. In one-fourth of the trials the
hand-to-cursor mapping is randomly off-set by several degrees. Users do
not expect these shifts and the analysis gives first insights that the
amplitudes of elicited ErrP could relate to the degree of error. If this
result is confirmed we may link error recognition to ``intuitivity''
evaluation.

\subsection{Emotions}

\subsubsection{Definition}

Psychology and neuroscience showed that emotions are connected to
high-level reasoning; they are tightly linked to decision-making
processes \citep{Damasio1994}. The valence/arousal model is the most
commonly used paradigm to categorize emotions \citep{Picard1995}. In
this two-dimensional representation, valence is related to hedonic tone
and varies from ``negative'' to ``positive'' (e.g.~frustrated vs
pleasant); arousal is related to bodily and mental activation and varies
from ``calm'' to ``excited'' (e.g.~satisfied vs happy). This model must
be applied with caution with some populations. Children hardly make
distinction between different arousal levels \citep{Posner2005}.

\subsubsection{Neuroimaging}

Technologies with the highest temporal resolution, such as MEG or EEG,
are more indicated when a dynamic content is involved
\citep{Vecchiato2011}.

An asymmetry within frequency bands (e.g.~alpha and theta) in the
frontal brain could be related to different emotions (valence), such as
pleasantness/unpleasantness \citep{Vecchiato2011}. Still, EEG is not yet
a reliable sensor to assess emotions. In \citep{Chanel2011} even if EEG
was better than the other studied physiological sensors on short period
of times, a 56\% accuracy barely suffices for the differentiation of
three emotions (chance level is 33\%).

Some Papers report high classifications rates. In \citep{Liu2011} 7
emotions are categorized. Authors state a 85\% accuracy for arousal and
90\% for valence. This using only three channels of an EEG headset which
is known to be sensitive to EMG artifacts. In pure EEG studies it is
important to control for facial expressions (i.e.~EMG signals), because
they can be easily recorder by electrodes. This is even more problematic
when emotions are involved. Although we have to be cautious when
assessing EEG reliability, there is nothing wrong in combining EEG and
EMG (or other sensors) to improve overall performance.

Despite the lack of clear indicators of affect in EEG, neuroimaging is
nevertheless a good lead for novel research in this topic. For example
different patterns of EEG signals have been observed depending on the
sense (sight or hearing) which induces an emotion \citep{Muhl2011}. It
could then be speculated that neuroimaging one day will be able to
discriminate which emotion is elicited by which input modality, or which
information channel leads to positive and which to negative user
experience.

\subsection{Engagement -- Flow -- Immersion}

\subsubsection{Definition}

Definitions of ``engagement'', ``immersion'' and ``flow'' overlap. From
\citep{Matthews2002}, task engagement is defined as an ``effortful
striving towards task goals''. Authors add that task engagement
increases during a demanding cognitive task and decreases when
participants perform a sustained and monotonous vigilance task, see also
\citep{Fairclough2009a}. In \citep{Chanel2011} ``engagement'' is treated
as one particular emotion, expressed as ``positive excited'' in the
valence/arousal model. Engagement is at a crossroads between several
concepts studied in this paper: workload, attention and emotions.

``Flow'' originates from psychological studies involving challenge
and/or creativity. It is a state in which someone is totally involved in
what he is doing. Flow happens when the skills of the person meet a
sufficient amount of challenge. A too important challenge brings
anxiety, for too much skills it is boredom, and too few of both results
in apathy \citep{Nacke2009}. Here again, several measures are involved.
Challenge relates to workload and the resulting state to emotions. By
definition, flow implies engagement.

``Immersion'' is studied mainly in virtual reality (VR) litterature. In
\citep{Slater2009} immersion stands for the modalities hardware gives to
users, how well devices can preserve fidelity in VR compared to reality.
Then the subjective feeling of being in the VR is called ``presence''.
Unfortunately this distinction between ``immersion'' and ``presence'' is
less clear-cut in other papers, see \citep{Nacke2009}.

\subsubsection{Neuroimaging}

In neuroimaging literature \citep{Fairclough2009a}, \citep{George2010}
engagement assessment studies are mentioned, but they often relate only
to sub-components such as workload or attention. \citep{Berka2007} see
engagement as a process related to information gathering, visual
scanning, and sustained attention. This study managed to discriminate
workload and engagement by using EEG but the tasks involved (mental
additions, recalls) are close to what is seen elsewhere in
attention/vigilance protocols. Engagement is often left entangled with
other states in a ``performance'' measure, see \citep[
sec.~3.2]{Blankertz2010}.

Experiments conducted during the FUGA project showed that flow could be
related to fMRI measures \citep{Ravaja2009}. The analysis with EEG of
frequency bands shows different pattern across three conditions of
interaction: boredom (i.e.~not engaged), flow and immersion in a pilot
study \citep{Nacke2010}. \citep{Berta2013} improved on this work and
achieved a 66\% classification accuracy.

\section{CHALLENGES}

\noindent We saw how constructs relevant to HCI can be investigated with
neuroimaging techniques. In this section we will argue that two of them
could benefit drastically from neurotechnologies: error recognition and
attention. Besides accuracy, both could reach a new level of
description. Furthermore we will emphasize the need for the evaluation
of a whole HCI to account for constructs of higher level, to study
usability and user experience. Finally we have to take care of EEG
devices and reliability in order to make it casual for experimenters to
use neuroimaging techniques.

\subsection{Improving on Constructs}

Measuring of two constructs would particularly benefit from improvements
in neuroimaging.

First, as it may enable a real-time measure of how intuitive a UI is, we
would benefit from a continuous and modulated measure of error
recognition. We saw how error recognition can be indicated through ErrP
\citep{Schalk2000}. This means that it is possible to detect when an
interaction runs against users' expectations \citep{Ferrez2008},
i.e.~when it is not intuitive. At the moment only a binary measure and
poorly detailed data -- ``an ErrP is detected or not'' -- is reliably
obtained. Fortunately it seems possible to measure a modulated ErrP
\citep{Sobolewski2013}, thus sensing by how much an operation in the UI
has perturbed users. If it is to be confirmed, this would enable a
quantitative and qualitative data assessment. We saw how single trial
detection can be achieved with EEG. Promising work reported ErrP
detection as the movement is occurring, within a 400ms timeframe
\citep{Milekovic2013}. At the moment this near continuous detection uses
an invasive technique.

The construct evolving around attention would be the second one to
profit from neuroimaging. To distinguish clearly in their measurements
vigilance and fatigue would be one point. On the other hand EEG studies
showed that visual artifacts in images or videos are detected by
subjects beyond consciousness \citep{Scholler2012}, whenever it is
conscious perception or attention \citep{Mustafa2012}. This would
suggest that ERP could be used to anticipate how much information users
are able to process, before even considering their attention level. A
(highly) speculative experimental design where various cues are hidden
within sensory modalities in order to elicit evoked potentials would
create a ``human bandwidth'' assessment, upstream from vigilance and
attention.

\subsection{Assessing New Constructs}

Three constructs sit apart in our nomenclature. Both usability and
comfort are more closely related to UI properties than to users' state,
and user experience is entirely based on previously seen measures. Since
they are the subject of many HCI papers, it is worth to shape their
meaning in this review in order to encourage their assessment with
neuroimaging techniques.

\subsubsection{Usability -- Comfort}

``Usability'' groups together the notions of ``ease of use'' and
``usefulness'' \citep{Bowman2002}. It relates to speed, accuracy and
error rates in task completion. The learnability of UI is also a key
point of usability. As such a good affordance of UI elements -- how
perceptions of objects induce a proper use -- will improve overall
usability. Usability is impacted by UI nature and constrains. E.g. an
input device based on body gestures is likely to be more tiring than a
joystick, given that it requires more energy from the user. Usability is
inextricably bound to users' comfort.

Although usability could be investigated through behavioral studies or
inquiries \citep{Jankowski2013}, to our knowledge there is no
neuroimaging study which accounts solely for this construct.
Neuroimaging has been used instead as an indicator, for example workload
through fNIRS \citep{Hirshfield2009}. In conjunction with other
evaluation methods, real-time recordings from physiological sensors and
neuroimaging give additional insights and help to contextualize data
\citep{Pikea2012}.

\subsubsection{User Experience}

For \citep{Mandryk2006} user experience (UX) is a shift from usability
analysis by bringing emotions and entertainment into the equation. UX
embeds ``usability/comfort'', ``emotions'' and
``engagement/flow/immersion''. UX is a higher comprehension level of
what users experience during interactions. \citep{Ravaja2009} compiled
various methods to measure media enjoyment. It is possible to refer to
UX when studying the social aspect of interactions -- e.g.~GSR is
different if the opponent in a sport game is played by a friend or a
computer \citep{Mandryk2006}. Assessing UX every time new technologies
are used could guide the HCI community in its choices, e.g.~with BCI
\citep{Laar2013}.

\subsection{Hardware -- Signal Processing}

Some limitations observed in EEG research are yet to be resolved to make
EEG-based evaluation of HCI more operable. EEG devices, while practical
compared to other neuroimaging techniques, take long to set up. Hence
experiments can be tedious both for the experimenter and for the
subject. This is why there are often only few subjects during EEG or BCI
experiments, which is a problem for the reliability of the results. EEG
signals contain many potential artifacts (e.g.~muscular activity and
electrical parasites); the quality of the device is essential. EEG
signals must be calibrated, processed and interpreted carefully.

Since a few years new EEG devices have appeared, oriented toward a
larger public. Their electrodes use no conductive solution, or water as
solvent. These electrodes are faster to set-up -- no more gel to be put
on each one after the device has been installed -- but may be less
sensitive, see \citep[ sec.~2.1]{Blankertz2010}. Hence some companies,
while transforming EEG into a mass-product, bring less reliable
technology to the market. Those devices often possess fewer electrodes.
Without a cap the electrodes are difficult to place in a standardized
position on the scalp. Finally they are often packaged with software
development kits which hide the signal processing from the users.
Constructs like attention or emotions are then claimed to be directly
measured, without further justification or muscular artifact control,
see \citep{Heingartner2009}. Nevertheless, while experimenters must be
aware of such limits if their intent is to rely solely on brain
activity, this increasing appeal in favor of cheap EEG devices is a
great opportunity to push forward the use of neuroimaging in HCI.

Improvements in signal processing, either in features extraction or
classification, could benefit every technology. Constructs, such as
emotions, are not yet accurately assessed with pure EEG signals. When
too many classes (e.g.~emotions and workload levels) are assessed
altogether, the classifier performance drops -- e.g.~see how the
``curse-of-dimensionality'' relates to classifiers' complexity
\citep{Friedman1997}. Improvements in mathematical analysis and machine
learning algorithms, as well as a better understanding of brain
activity, would increase the reliability of the whole system by a great
amount and favour every construct.

Finally, no matter how lightweight they are, EEG and physiological
sensors change the way users interact. Movements could be restrained by
the devices (less immersion) and users could perceive a more stressful
context, potentially biasing their experience. As a result, a framework
integrating physiological sensors and traditional evaluation methods has
to be conceived to profit from the potential of these novel methods,
while avoiding their limitations and pitfalls.

\section{CONCLUSION}

\noindent We reviewed how neuroimaging techniques could assess
constructs relevant for HCI evaluation.

Between the four categories of evaluation methods, inquiries could
deliver more qualitative data, while physiological sensors and
neuroimaging are exocentric measures (the most ``objective'' measures of
subjectively perceived stimuli). It is particularly interesting to
combine those methods for constructs otherwise difficult to assess with
exactitude, as investigated in many studies \citep{Ravaja2009},
\citep{Nacke2009}, \citep{Erp2010}, \citep{Chanel2011}.

Our analysis of neuroimaging techniques focused on EEG as it promises a
good trade-off between cost, time resolution and ease of installation.
We advocate that neurotechnologies can bring useful insights to HCI
evaluation. EEG devices are not yet perfectly reliable and practical to
use; hardware and software processing are still evolving. However their
cumbersomeness is partially avoided if they are used during a dedicated
evaluation phase in the HCI development process, with specially enrolled
users (testers).

\begin{figure}[htbp]
\centering
\includegraphics{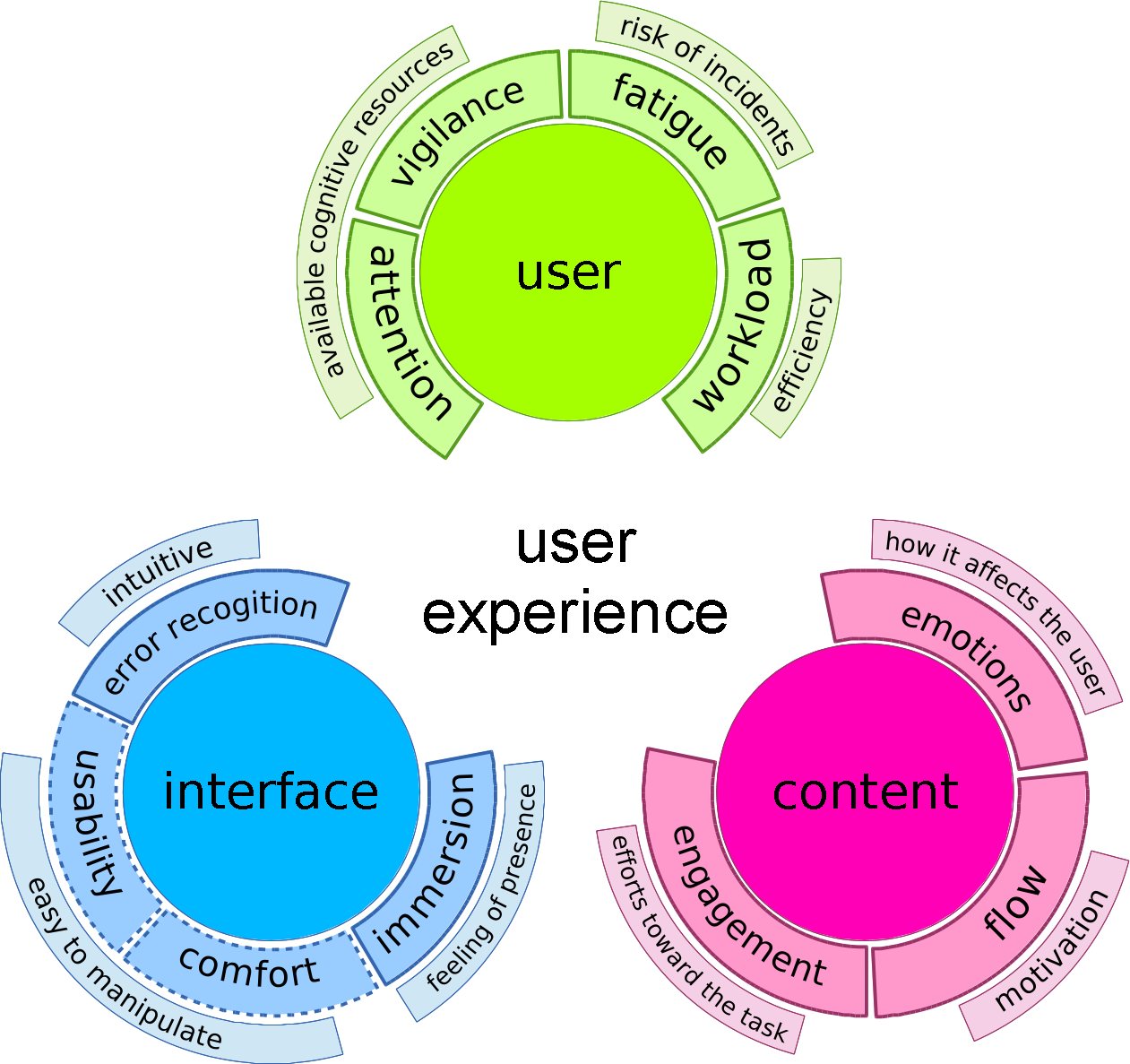}
\caption{One possible view of a simplified characterization of the
constructs. In the middle circles are the constructs (dotted = not yet
measurable with EEG). The inner circles represent the HCI components the
most closely related to the constructs, or on which it would be easier
to leverage. The outer circles give a hint about what an evaluation
would be useful for. \label{constructs_simple}}
\end{figure}

We studied workload, attention, vigilance, fatigue, error recognition,
emotions, engagement, flow and immersion. Figure \ref{constructs_simple}
stimulates thoughts about their relationships with HCI components. Some
constructs should benefit more than the others from EEG measures: 1)
workload, EEG being more sensible to changes compared to other methods
\citep{Mathan2007}; 2) attention, because event related potentials could
help to anticipate how many details users register \citep{Mustafa2012};
3) emotions, with an arousal/valence state measured over a short
time-frame \citep{Chanel2011}. Error recognition could hardly be
assessed precisely with anything but neuroimaging. Such construct
highlights how innovative this evaluation method is. Among the outlined
challenges, a continuous and modulated error recognition would greatly
help to assess usability and comfort.

Next studies should start to combine the various constructs, along with
a comprehensive framework which gathers every evaluation method, one's
advantages preventing others' drawbacks. This should lead to an increase
of the overall user experience.

\bibliographystyle{apalike}
{\small
\bibliography{biblio}}

\end{document}